\documentclass[aps,pr,twocolumn,showpacs,superscriptaddress,groupedaddress]{revtex4}  

\usepackage{graphicx}  
\usepackage{dcolumn}   
\usepackage{bm}        
\usepackage{amsmath}
\usepackage{amssymb}   
\usepackage{epsfig}

\begin{document}
\widetext
\leftline{\hspace{5 in}FERMILAB-PUB-12-251-PPD}
\vspace{0.25 in}

%
\def \ie    {\hbox{\it i.e.}}     
\def \etc   {\hbox{\it etc.}}
\def \ibid  {\hbox{\it ibid.}}
\def \vs    {\hbox{\it vs.}}
\def \eg    {\hbox{\it e.g.}}     
\def \cf    {\hbox{\it cf.}}
\def \etal  {\hbox{\it et al.}}
\def \via   {\hbox{\it via}}
\def \CPC   {Comput. Phys. Commun.~}
\def \EPJ   {European Phys. Journal~}
\def \JHEP  {J. High Energy Phys.~}
\def \NIM   {Nucl. Instr. Meth.~}
\def \NP    {Nucl. Phys.~}
\def \PL    {Phys. Lett.~}
\def \PRD   {Phys. Rev. D~}
\def \PRL   {Phys. Rev. Lett.~}
\def \PRep  {Phys. Reports~}
\def \RMP   {Rev. Mod. Phys.~}
\def \zphys {Z. Phys.~}
\def \JPhG  {J. Phys. G: Nucl. Part. Phys.~}
\hyphenation{back-ground}
\hyphenation{brem-sstrah-lung}
\hyphenation{cal-or-ime-ter cal-or-ime-try}
\hyphenation{had-ron had-ronic}
\hyphenation{like-li-hood}
\hyphenation{posi-tron posi-trons}
\hyphenation{semi-lep-tonic}
\hyphenation{syn-chro-tron}
\hyphenation{system-atic}
%
%
\def \beq   {\begin{equation}}
\def \eeq   {\end{equation}}
\def \bbeq  {\begin{eqnarray*}}
\def \ebeq  {\end{eqnarray*}}
\def \bbeqn  {\begin{eqnarray}}
\def \ebeqn  {\end{eqnarray}}
\def \Tr    {\mathop{\mathrm Tr}}
\def \Im    {\mathop{\mathrm Im}}
\def \Re    {\mathop{\mathrm Re}}
\def \vect  {\overrightarrow}
\def \twdl  {\widetilde}
\def \hat   {\widehat}
\def \partder#1#2  {\partial #1\over\partial #2}
\def \secder#1#2#3 {\partial^2 #1\over\partial #2 \partial #3}
%
%
\def \omg#1 {\mbox {${\mathcal O}(#1)$}}
\def \avg#1 {$\left\langle #1\right\rangle$}
\def \to    {\rightarrow}
\def \bra#1 {$\left\langle #1\right|$}
\def \ket#1 {$\left| #1\right\rangle$}
\def \braket#1#2 {\left\langle #1\right. \left| #2\right\rangle}
\def \amp#1 {${\mathcalA}(#1)$}
\def \apgt  {\raisebox{-0.6ex}{$\stackrel{>}{\sim}$}}
\def \aplt  {\raisebox{-0.6ex}{$\stackrel{<}{\sim}$}}
\def \pma#1#2 {\mbox{\raisebox{-0.6ex}
           {$\stackrel{\scriptstyle \;+\; #1}{\scriptstyle \;-\; #2}$}}}
%
\def \dr    {$\;^\mid\!\!\!\longrightarrow$}
%
\def \ev    {\,\mathrm {eV}}
\def \kev   {\,\mathrm {keV}}
\def \mev   {\,\mathrm {MeV}}
\def \gev   {\,\mathrm {GeV}}
\def \tev   {\,\mathrm {TeV}}
\def \km    {\,\mathrm {km}}
\def \cm    {\,\mathrm {cm}}
\def \mm    {\,\mathrm {mm}}
\def \um    {\,\mu\mathrm m}
\def \ghz   {\,\mathrm {GHz}}
\def \mhz   {\,\mathrm {MHz}}
\def \khz   {\,\mathrm {kHz}}
\def \ps    {\,\mathrm {ps}}
\def \ns    {\,\mathrm {ns}}
\def \us    {\,\mu\mathrm s}
\def \ms    {\,\mathrm {ms}}
\def \hz    {\,\mathrm {Hz}}
\def \pb    {\,\mathrm {pb}^{-1}}
\def \fb    {\,\mathrm {fb}^{-1}}
\def \mrad  {\,\mathrm {mrad}}
\def \BR#1#2 {\mbox{Br}(#1$\to$#2)}
\def \JP     {\mathrm J$^{\mathrm P}$}
\def \Mw    {${\mathrm M}_W$}
\def \Mz    {${\mathrm M}_Z$}
\def \Mpi   {${\mathrm M}_\pi$}
\def \Mk    {${\mathrm M}_K$}
\def \Gf    {G$_{\mathrm F}$}
\def \As    {$\alpha_s$}
\def \Mt    {${\mathrm M}_{t}$}
\def \Mb    {${\mathrm M}_{b}$}
\def \Mc    {${\mathrm M}_{c}$}
\def \Ms    {${\mathrm M}_{s}$}
\def \Mud   {${\mathrm M}_{u,d}$}
\def \Mh    {${\mathrm M}_{H}$}
\def \sW    {$\sin^2\theta_{\mathrm W}$}
\def \sWeff {$\sin^2\theta_{\mathrm W}^{eff}$}
\def \gV    {g$_V$}
\def \gA    {g$_A$}
\def \lms   {\Lambda_{\overline{\mathrm MS}}}
\def \Vub   {$|$V$_{\mathrm{ub}}|$}
\def \Vus   {$|$V$_{\mathrm{us}}|$}
\def \Vcb   {$|$V$_{\mathrm{cb}}|$}
\def \MET   {\mbox{$E_T\hspace{-1.2em}\slash\hspace{1.0em}$}}
\def \PET   {\mbox{$p_T$}}
\def \pT    {\mbox{$p_\perp}}
\def \dedx {d{\it E}/d{\it x}}
\def \rphi {$r$-$\phi$}
\def \Pt2  {${\mathrm P}_\perp^2$}
%
\def \epem {$e^+e^-$}
\def \mpmm {$\mu^+\mu^-$}
\def \tptm {$\tau^+\tau^-$}
\def \ppb  {$p\overline p$}
\def \bbb  {$b\overline b$}
\def \ttb  {$t\overline t$}
\def \lplm {$\ell^+ \ell^- $}
\def \J    {$J/\psi$}
\def \Ks   {$K^0_{\mathrm S}$}
\def \Kl   {$K^0_{\mathrm L}$}
\def \Bs   {$B_{\mathrm S}$}
\def \Bo   {$B^0$}
\def \Bp   {$B^+$}
\def \Ds   {$D_{\mathrm S}$}
\def \Do   {$D^0$}
\def \Dp   {$D^+$}
\def \Bss  {$B_{\mathrm S}^*$}
\def \Bso  {$B^{*0}$}
\def \Bsp  {$B^{*+}$}
\def \Dss  {$D_{\mathrm S}^*$}
\def \Dso  {$D^{*0}$}
\def \Dsp  {$D^{*+}$}

\def \mN {$m(\nu_4)$~}
\def \mE {$m(\tau_4)$~}
\def \mT {$m(t_4)$~}
\def \mB {$m(b_4)$~}
\def \MC {{\sc 2HDMC}}
\def \HB {H{\sc iggs}B{\sc ounds}}
\def \tanB {$\tan \beta$}

\title{Masses of a Fourth Generation with Two Higgs Doublets}

\author{Leo~Bellantoni}
	\email{bellanto@fnal.gov}
	\affiliation{Fermi National Accelerator Laboratory, Batavia, IL 60510, USA}
\author{Jens~Erler}
	\email{erler@fisica.unam.mx}
	\affiliation{Departamento de F\'isica Te\'orica, Instituto de F\'isica,
	 Universidad Nacional Aut\'onoma de M\'exico, 04510 M\'exico D.F., M\'exico}
\author{Jonathan~J.~Heckman}
	\email{jheckman@ias.edu}
	\affiliation{School of Natural Sciences, Institute for Advanced Study,
	Princeton, NJ 08540, USA }
\author{Enrique~Ramirez-Homs}
	\email{enraho@gmail.com}
	\affiliation{University of Texas, El Paso, TX 79968, USA}

\date{21 June 2012}

\begin{abstract}
We use sampling techniques to find robust constraints on the masses of a
possible fourth sequential fermion generation from electroweak oblique
variables.  We find that in the case of a light ($115\gev$) Higgs from
a single electroweak symmetry breaking doublet, inverted mass hierarchies
are possible for both quarks and leptons, but a mass splitting more than
$M_W$ in the quark sector is unlikely.  We also find constraints in the
case of a heavy ($600\gev$) Higgs in a single doublet model.  As recent
data from the Large Hadron Collider hints at the existence of a resonance
at $124.5\gev$ and a single Higgs doublet at that mass is inconsistent
with a fourth fermion generation, we examine a type II two Higgs doublet
model.  In this model, there are ranges of parameter space where the
Higgs sector can potentially counteract the effects of the
fourth generation. Even so, we find that such scenarios produce
qualitatively similar fermion mass distributions.

\end{abstract}

\maketitle

\section{\label{sec:intro}INTRODUCTION}

Adding a sequential fourth generation of fermions (4G) is one of the
simplest possible extensions to the Standard Model. Indeed, although
the width of the $Z$ limits the number of active light neutrinos to three, there
can in principle be a fourth neutrino generation which is much heavier.
A recent and extensive literature examines the impact of how 4G would reduce
tensions in recent measurements in the $b$ sector and create distinctive
phenomena in kaon decays~\cite{bib:SoniEtTu,bib:BurasEtal}, as well as providing a potential scenario for
the baryon asymmetry of the universe~\cite{bib:WS_Hou}. As a more top down
motivation, simple string constructions often
lead to toy models with an even number of generations. Of course, achieving
three chiral generations is also possible in a wide class of examples, and
some stringy models of flavor physics predict that more than three generations
would be inconsistent with the measured three generation quark
mixing matrix~\cite{bib:Heckman:2008qa}. More generally, one can view the 4G
scenario as a simple template for scenarios of physics beyond the Standard
Model in which states of some extra sector receive a mass proportional to the
Higgs vev.

In light of these considerations, it is clearly of interest to study the
viability of the 4G scenario.  In addition to the possibility of direct
detection of such states, the contributions of these additional states
enter as loop corrections to various Standard Model processes. For example, a
fourth generation tends to produce a positive contribution to both $S$ and
$T$, the oblique electroweak parameters. By contrast, in a single Higgs doublet
model, increasing the mass of the Higgs generates a positive contribution to
$S$ and a negative contribution to $T$.  Thus, while cancellation for the $T$
parameter is possible, the contributions to the $S$ parameter typically move
in the same (positive) direction, though mass hierarchies in the fourth
generation can reduce the size of this contribution. The extra generation also
affects the phenomenology of the Higgs, leading to an increase in
$\Gamma(h \rightarrow g g)$, and a decrease in
$\Gamma(h \rightarrow \gamma \gamma)$.

Though less well-studied, even simple extensions of the Higgs sector can
counteract (or exacerbate) some of the effects of a chiral fourth generation
with an appropriate tuning of parameters. For example, in two Higgs doublet
models (2HDM), the contributions to $S$ and $T$ can have either
sign (see e.g. \cite{bib:Heckman:2011bb,bib:Funk:2011ad,bib:HaberONeil}).
General values of the Higgs mixing angles also allow for changes in
$\Gamma(h \rightarrow g g)$ and $\Gamma(h \rightarrow \gamma \gamma)$,
independently, relative to the 4G scenario.

\begin{figure*}[t]
\begin{center}
\includegraphics[width=\textwidth]{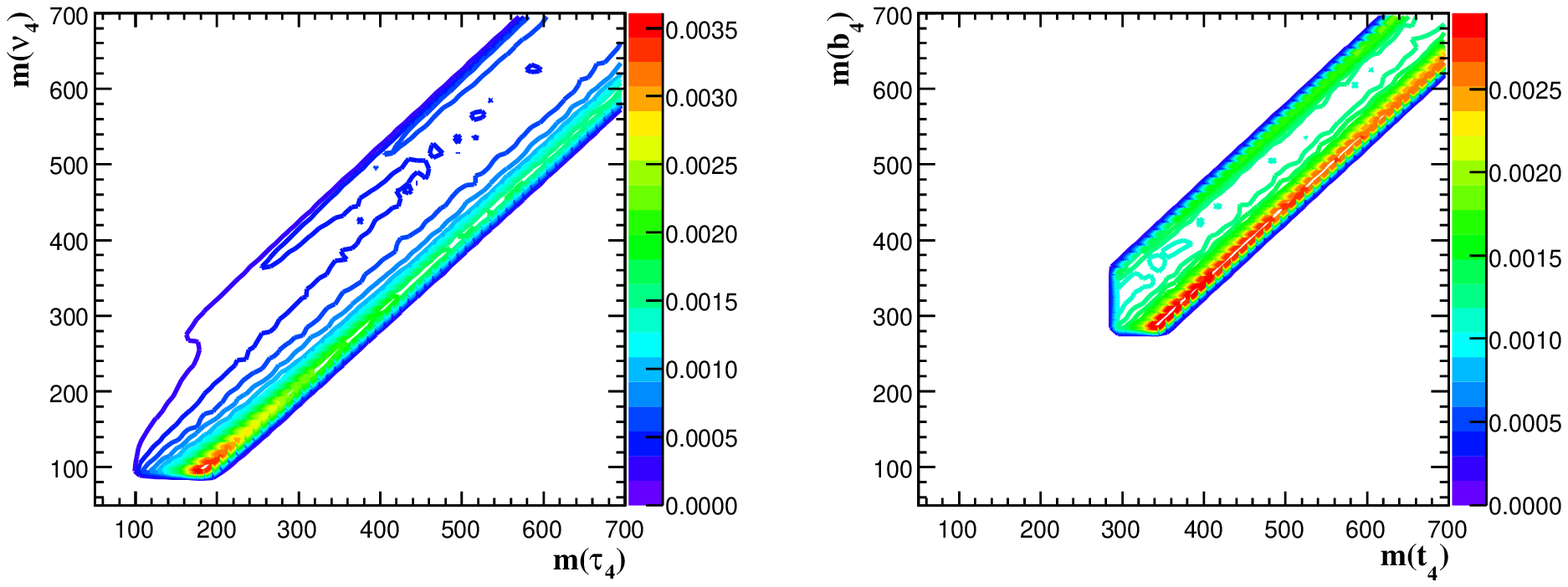}
\caption{Contour plots of the probability densities in the 4G
baseline scenario. Left: \mN \vs~\mE;   Right: \mB \vs~\mT;  All scales are in
$\gev$; probability densities have been normalized and each bin is
$10\gev \times 10\gev$.}
\label{fig:base_M12}
\end{center}
\vspace{0.2em}
\end{figure*}

\begin{figure*}[t]
\begin{center}
\includegraphics[width=\textwidth]{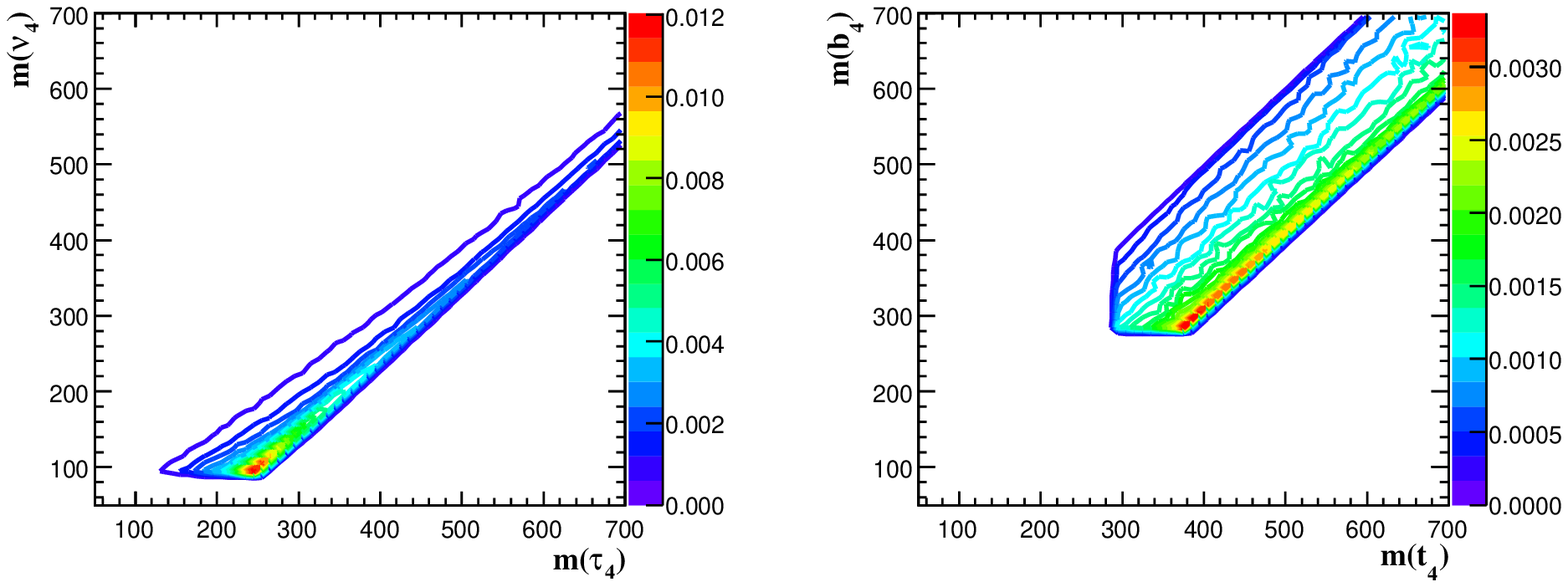}
\caption{Contour plots of the probability densities in the 4G
scenario with $m(h) = 600$ GeV.  Left: \mN \vs~\mE;   Right: \mB \vs~\mT;  All
scales are in $\gev$; probability densities have been normalized and each bin
is $10\gev \times 10\gev$.}
\label{fig:high_M12}
\end{center}
\vspace{0.2em}
\end{figure*}

\begin{figure*}[t]
\begin{center}
\includegraphics[width=0.45\textwidth]{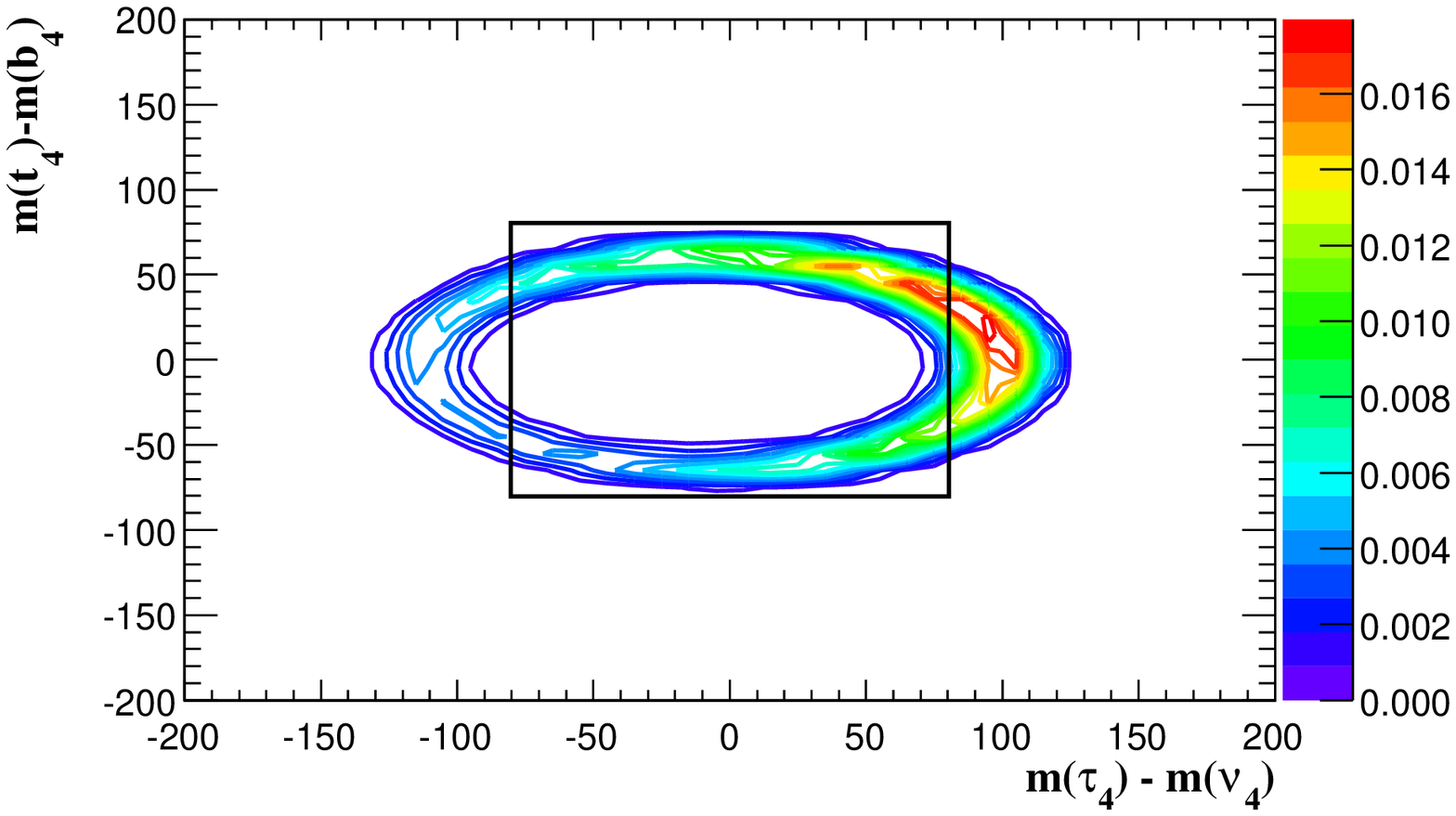}
\hspace{0.05\textwidth}
\includegraphics[width=0.45\textwidth]{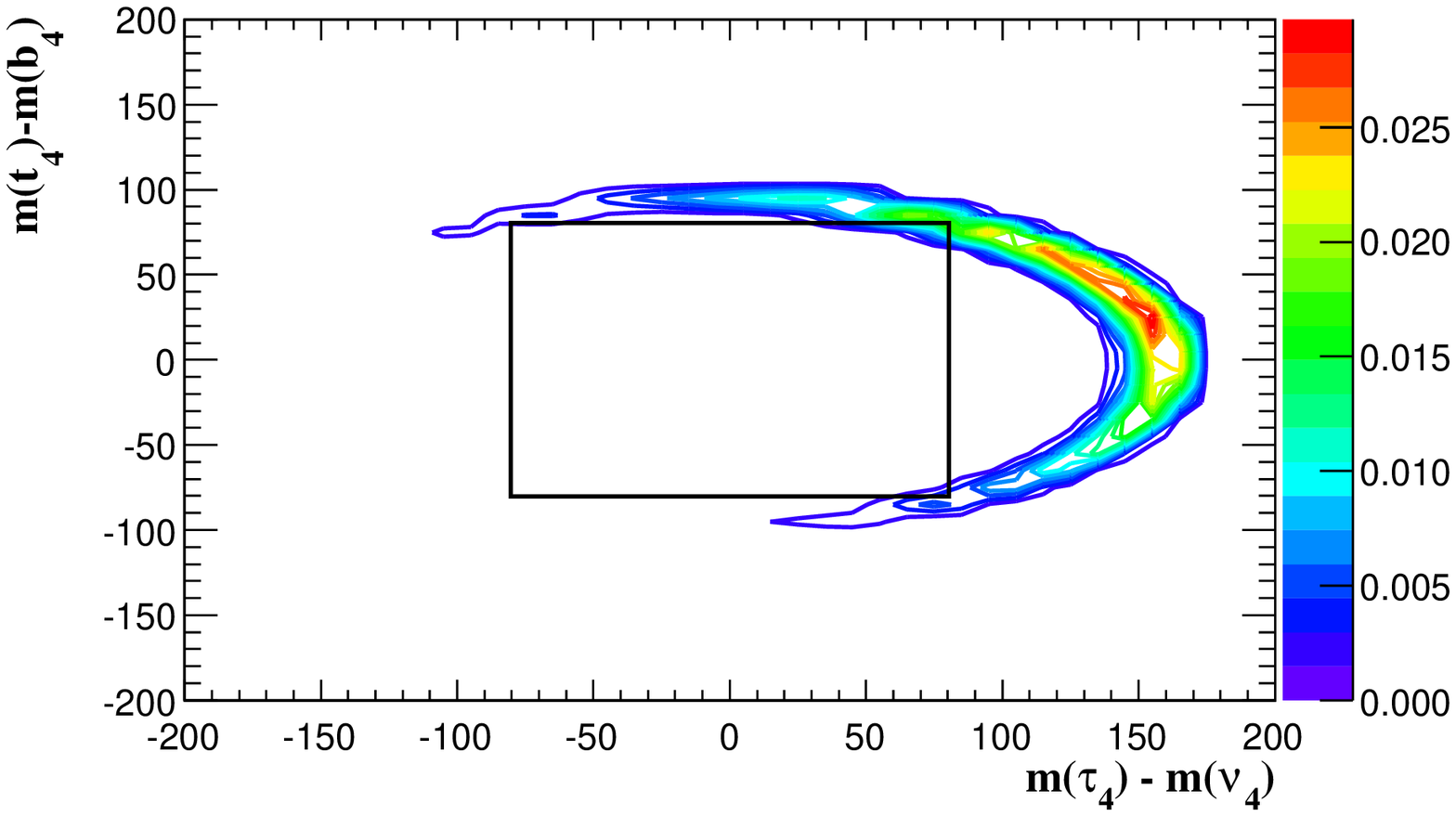}
\caption{Contour plots of the probability densities for quark \vs~lepton mass
splitting in the 4G scenario.   Left: with $m(h) = 115\gev$;   Right: with
$m(h) = 600\gev$.  The boxes mark the areas where the magnitude of the mass
splittings is less than $M_W$.  All scales are in $\gev$; probability
densities have been normalized and each bin is $10\gev \times 10\gev$.}
\label{fig:base_high_dM}
\end{center}
\vspace{0.2em}
\end{figure*}

In this paper we study the available parameter space for the 4G scenario
and its extension to 2HDM models. The full parameter space of 4G is too large for
easy visualization, but much of what we need to know to understand existing
experimental results and to inform future searches can be expressed with two
pairs of numbers: the two quark masses and the two lepton masses.  We
therefore seek, by sampling this four-parameter space and comparing the samples
with constraints on electroweak oblique parameters to determine the most
likely mass spectrum for 4G, should it exist. Similar earlier analyses
of this type may be found in~\cite{bib:HePoSu,bib:Tim,bib:GFitter}, but
new experimental data has appeared since these publications.  Other
similar studies have appeared recently~\cite{bib:GoodInj,bib:CloseCall,bib:A_VV}.

Exclusion limits on the mass of a Standard Model-like Higgs impose additional
constraints on the 4G scenario. In 4G, the gluon fusion production cross section
for the Higgs is markedly increased over the three generation scenario.  Both
LHC collaborations~\cite{bib:LHC_Higgs4} independently exclude, using a
combination of channels, the range $120\gev < M_{h} < 600\gev$~ when there is
a fourth generation.  The LEP II lower limit of $114.4\gev$ is independent of
the number of fermion generations~\cite{bib:LEP_Higgs}.  Because a fourth
generation of fermions contributes to $T$ roughly quadratically with \mT and
$m(b_4)$, and because a large $T$ corresponds to a large $m(h)$, values of
$m(h)$ as high as $1\tev$ are allowed by electroweak
constraints~\cite{bib:TwoFits,bib:GFitter} in 4G.  However, studies of the
stability and triviality bounds on $m(h)$ in 4G~\cite{bib:MichioAkin} prohibit
$m(h) \gtrsim 700\gev$ unless there is also some other new phenomena on a scale
below $2\tev$.

Most recently, there are ``hints'' of a Higgs with mass~\cite{bib:JensWeighs}
of $124.5 \pm 0.8 \gev$ from the LHC~\cite{bib:LittleBump1} and supporting
evidence from the Tevatron~\cite{bib:LittleBump2}.  The hint is strongest
in the channel $g \, g \rightarrow h \to \gamma\gamma$, where the ATLAS
experiment reports an excess above background of 2.8 standard deviations.
The statistical significance of these results is not enough to declare
discovery or even strong evidence for a Higgs, but is strong enough to provoke
discussion.  This mass is within the bounds ruled out by the LHC when supposing
a fourth fermion generation.  To leading ${\mathcal O}(G_F m^2_f)$, it
is possible to retain 4G if one supposes only the $\gamma\gamma$~ channel's
hint remains significant with the addition of more data, but including exact
next-to-leading order electroweak corrections makes this
difficult~\cite{bib:gamgamOK1,bib:gamgamOK2}.  Consequently, the 4G
hypothesis is valid only if (a) the hints turn out to be statistical
fluctuations or (b) the hints are due to something beyond a single Higgs
doublet, such as a two Higgs doublet model.

The rest of this paper is organized as follows. In Section~\ref{sec:singles} we treat the 4G
single Higgs doublet case. We give results for
$m(h) = 115$ GeV (the ``baseline'' scenario) and for $m(h) = 600$ GeV (the
``high-mass'' scenario).  The baseline scenario is appropriate for the case
considered in~\cite{bib:gamgamOK1,bib:gamgamOK2}; the difference in our results between
$m(h) = 115$ GeV and $124.5$ GeV is small. In Section~\ref{sec:doubles}
we extend the analysis to consider a Type II
model with its parameters adjusted to match the $124.5\gev$ hint.
Section~\ref{sec:sum} provides a summary.

\section{\label{sec:singles}SINGLE HIGGS DOUBLET SCENARIOS}

\subsection{Method}

We have updated constraints on the oblique electroweak~\cite{bib:PandT}
parameters $S$, $T$ and $U$, as found by the Global Analysis of Particle
Properties (GAPP)~\cite{bib:GAPP} using data available in October 2011.  In
our sampling procedure, each sample is assigned a weight corresponding to
the probability density function $p = p(S,T,U)$ for these three parameters.
We employ the one-loop contributions to the oblique parameters, assuming small
mixing with the extra family, as in~\cite{bib:HePoSu}.
See~\cite{bib:Eberhardt:2012sb} for some recent discussion of the more general
case of potentially large mixing effects.

The sampling distribution in this type of analysis plays the role of a
Bayesian prior; we are taking the probability $p$ of a specific value for
$S$, $T$ and $U$ given an assumed set of four fermion masses, and
weighting it in our result as the probability density created by our
sampling of the fermion spectrum.  We interpret the result as a
probability density function for the fermion mass spectrum, but that
interpretation is only valid in the context of that assumed sampling
distribution.  The peril in this process - the validity of the assumed
prior - thus has the advantage of requiring explicit description.


We draw 50 million uniformly distributed samples in the fermion mass
spectrum with lower bounds set by direct experimental constraints
described below.  The upper bound is limited by unitarity
arguments~\cite{bib:unitarity} to $500\gev$, but this is a rough bound
and we raise it to $700\gev$ for clarity in the resulting figures.


The lower bound on the sampled \mN mass range is, in our baseline
scenario, \mN = $90.3\gev$ from LEP II~\cite{bib:L3}.  This limit is the
weakest of the limits obtained under the assumption of $\nu_4$ decay to
each of the three known charged leptons; if $m(\nu_{4}) > m(\tau_{4})$,
then we would obtain a stronger limit. The lower bound on the sampled \mE range is $100.8\gev$;
again, this is the weakest limit obtained in all the possible decay
scenarios.  These results are therefore robust against all assumptions
about the lepton mass hierarchies. On the other hand, lepton mixing
parameters are important considerations in searches for the leptons of
4G at the LHC which have been discussed~\cite{bib:4thG_L} but have not
yet been carried out.

Obtaining robust lower bounds on 4G quark masses and mixing angles is a little
more complicated. Dramatic results~\footnote{All limits are at the 95\%
confidence level} from the LHC are indeed available~\cite{bib:PIConf}, and
new ones are appearing constantly. The CMS collaboration has searched for:

\begin{itemize}

\item  $b_4{\overline {b_4}} \to tW^- {\overline t}W^+ \to
(bW^+W^-)({\overline b}W^-W^+)$ with same-charge leptons and trileptons in
a $4.6\fb$ sample~\cite{bib:CMS_p}, obtaining a limit of $600\gev$.

\item both $t_4$ and $b_4$ using a simplified model with a range of
final states, all containing 2 $b$ quarks, in $1.1\fb$ of
data~\cite{bib:CMS_2}.  All of the diagrams considered have $b_4 \to tW$
or $t_4 \to bW$.  Lower limits of $480-540\gev$ were obtained.

\item pair produced $t_4$ in the ``lepton with jets'' channel, wherein a
decay to $bW$ having the same signature as a $t{\overline t}$ event but
with a different primary quark mass is sought.  The analysis reconstructed
\mT in each event.  A $560\gev$ lower limit was found using only
$4.6\fb$ of data~\cite{bib:CMS_3}.

\item pair produced $t_4$ in the ``dilepton'' channel, wherein also a decay
to $bW$ having a top-quark signature but different mass is sought.  A weaker
constraint than that which was obtained in the ``lepton with jets''
analysis, $422\gev$, was found using $1.1\fb$ of data~\cite{bib:CMS_4}.

\end{itemize}

The ATLAS collaboration has searched for:

\begin{itemize}

\item pair produced $b_4 \to tW$
in $34\pb$ of data~\cite{bib:ATLAS_1}, as part of an inclusive search for
exotic production of the same-charge dilepton signature.

\item pair produced $t_4$ or $b_4$ decaying to $Wq$, where $q = u, d, s,$
or $b$, appearing with opposite-charge dileptons and missing transverse
momentum in $37\pb$ of data~\cite{bib:ATLAS_2}.  An approximate event
reconstruction is done.  The resulting limit is \mT = \mB $> 270\gev$.

\item  $b_4{\overline {b_4}} \to tW^- {\overline t}W^+ \to
(bW^+W^-)({\overline b}W^-W^+)$ with one lepton, at least six jets, and
large missing momentum transverse to the beamline on a $1.0\fb$ sample,
obtaining~\cite{bib:ATLAS_3} a limit of $480\gev$.

\item pair produced $t_4$ or $b_4$ appearing with same-charge dileptons,
large missing transverse momentum, and at least two jets in $1.0\fb$ of
data~\cite{bib:ATLAS_4}.  A limit of \mB $> 450\gev$~ was obtained.

\end{itemize}
See ~\cite{bib:wierdLHC} for recent searches of more exotic fermions.

These search results, while impressive, are all built upon specific decay,
\ie~CKM mixing angle, assumptions. With the exception
of~\cite{bib:ATLAS_2}, mixing of the fourth generation into anything other
than the third generation is not considered.  Furthermore,
$t_4 \to b_4 W^*$ (or in an inverted hierarchy, $b_4 \to t_4 W^*$) will be
an additional contribution to $b_4$ (or $t_4$) production which will not
necessarily appear in any specific signature as a result of the $W^*$
products; the contribution from this channel can be significant if the
mass splitting is small.

\begin{figure*}[t]
\begin{center}
\includegraphics[width=\textwidth]{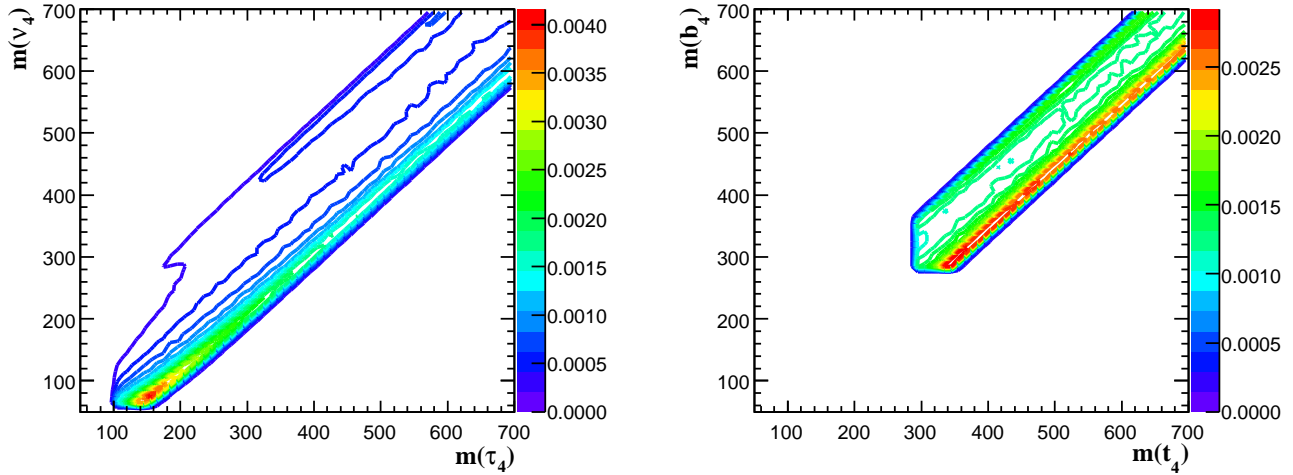}
\caption{Contour plots of the probability densities in the 4G baseline when
\mN is allowed to go as low as $60.1\gev$. Left: \mN \vs~\mE;   Right:
\mB \vs~\mT;  All scales are in $\gev$; probability densities have been
normalized and each bin is $10\gev \times 10\gev$.}
\label{fig:RHn4_M12}
\end{center}
\vspace{0.2em}
\end{figure*}

Additionally, there are constraints on the possible mixing parameters.
For example, the mixing parameters for the quark sector may be
constrained~\cite{bib:BurasEtal,bib:Mixing1} with data from neutral mesons,
the $b \to s\gamma$ transition, existing constraints on the three-generation
quark mixing matrix and limits on \BR{\Bs~}{\mpmm} .
Reference~\cite{bib:Mixing1} concludes that large mixings of the fourth
generation with the three known generations are not ruled out,
but~\cite{bib:Mixing2,bib:Mixing3}, which considers constraints from corrections
to the $Z \to$\bbb~ vertex from a fourth generation conclude that these
mixings could be comparable to Cabibbo mixing.  The quark mixing matrix can also
be constrained with precision electroweak data and
$D^0- \overline{D^0}$ mixing~\cite{bib:Mixing4}.  In any case however,
there is the possibility that 4G fermions could decay to either third or
lower generation fermions with varying branching ratios.

A method for producing experimental limits that are mixing-angle
independent~\footnote{P.Q. Hung and M. Sher, \PRD {\bf 77}, 037302 (2008)
point out that for very small mixing angles, 4G quarks are charged massive
particles, with signatures very different from those typically used in 4G
searches.} and that allows for the contributions of both 4G quarks to any
particular signature was applied to the results of CDF
searches~\cite{bib:Irvine}, resulting in lower limits of $\sim 280\gev$ for
\mB and $\sim 290\gev$ for $m(\tau_{4})$. We use these lower but mixing
independent values here while strongly advocating the application of these
techniques to the more recent LHC results.  Such an analysis could soon
sharply constrain or even rule out the 4G hypothesis.

\subsection{Results}

Figures~\ref{fig:base_M12} and~\ref{fig:high_M12} show the lepton and quark
mass spectra in our baseline and high $m(h)$~scenarios.  
In these and similar Figures, the color for each bin  represents a
probability density integrated over the bin, and normalized so as to give
unit probability when summed over the entire plot.  For the baseline
(high mass) case, $|m(t_4) - m(b_4)| < M_W$ in over 99\% (90\%) of our
samples; transitions between 4G quarks will produce off-shell $W$~bosons.
The lepton mass splitting is less than $M_W$ with probability 69\% (24\%).
Normal mass hierarchies are more likely than not, but by no means certain;
in the lepton sector the probability of a normal mass hierarchy is 70\% (93\%),
and in the quark sector, it is 59\% (69\%).

Masses just over the existing limit for the leptons are heavily favored, and
this tendency is greater in the high $m(h)$~scenario.  Being able to predict
this parameter relatively precisely makes it a valuable target for future
searches.

In Figure~\ref{fig:base_high_dM} we show the lepton and quark mass splittings.
We see that, with perhaps a two-fold ambiguity, the mass splittings in the
two sectors are tightly related.

\begin{figure*}[t]
\begin{center}
\includegraphics[width=\textwidth]{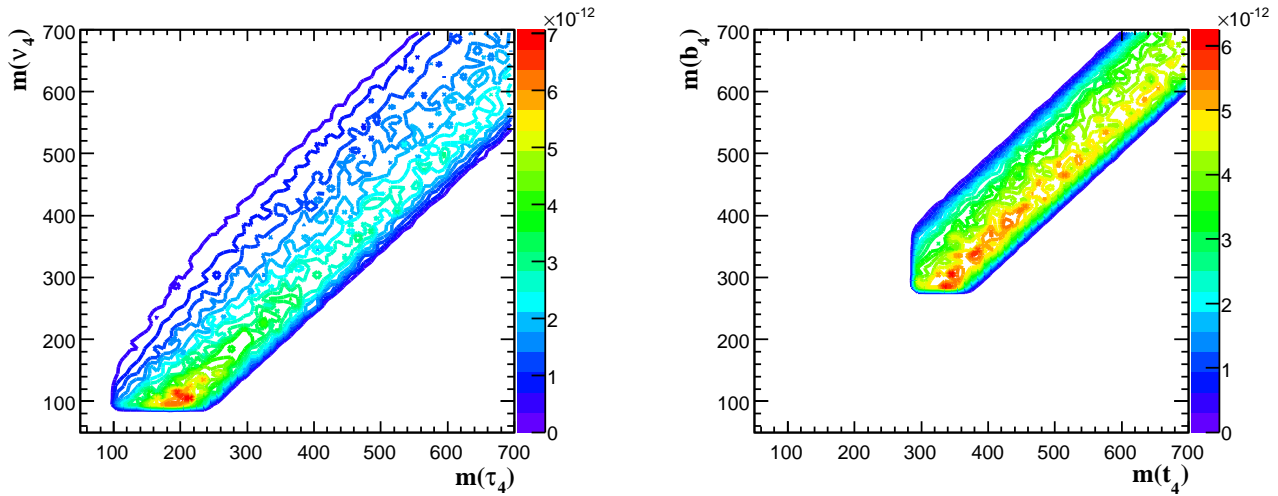}
\caption{Contour plots of the probability densities in 2HD4G with the mass of
the lightest ${\mathcal CP-}$even state $m(h) = 124.5$ GeV. Left: \mN \vs~\mE;
Right: \mB \vs~\mT;  All scales are in $\gev$; probability densities have been
normalized and each bin is $10\gev \times 10\gev$.}
\label{fig:THDM_M12}
\end{center}
\vspace{0.2em}
\end{figure*}

\begin{figure*}[t]
\begin{center}
\includegraphics[width=0.45\textwidth]{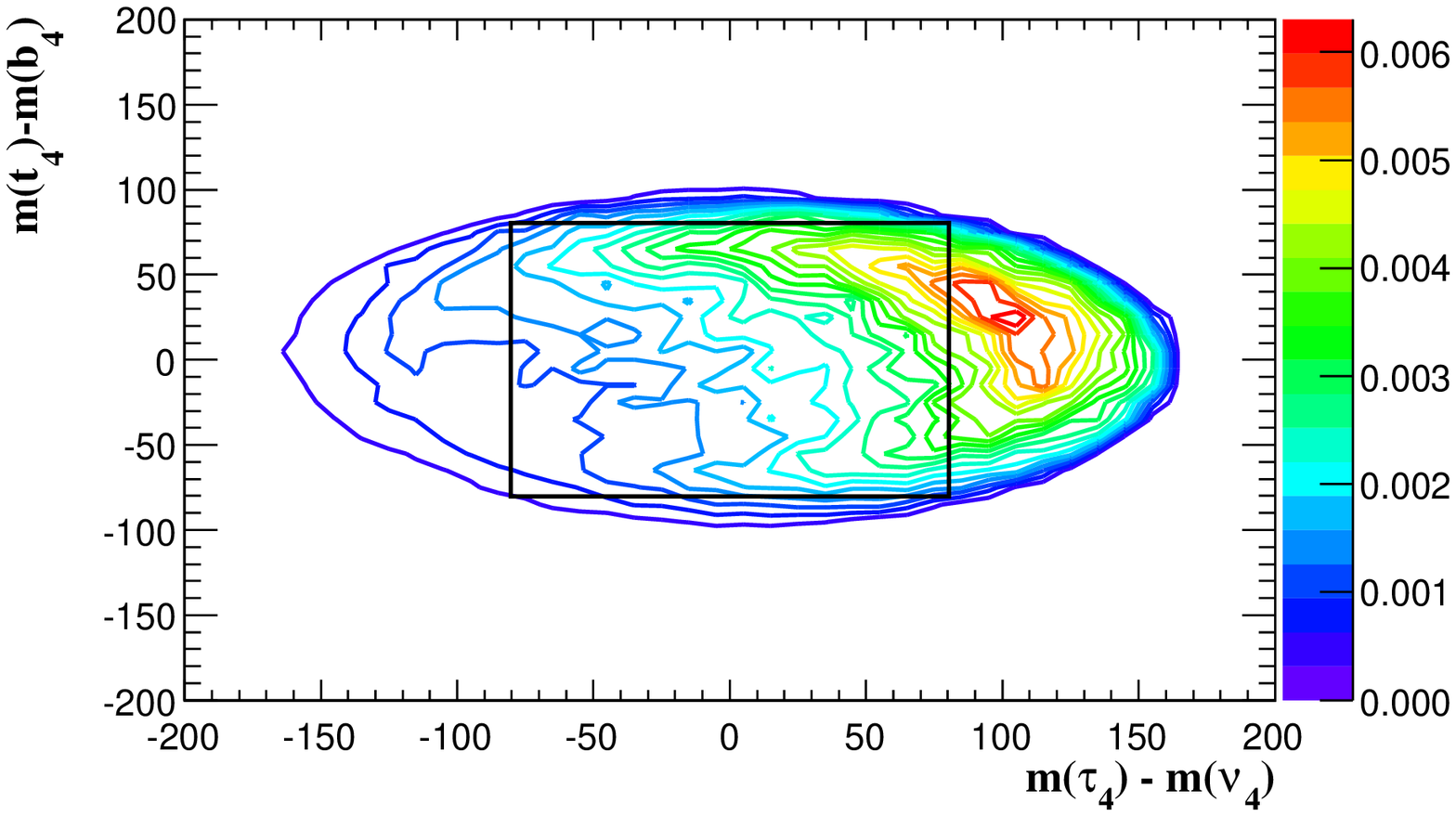}
\hspace{0.05\textwidth}
\includegraphics[width=0.45\textwidth]{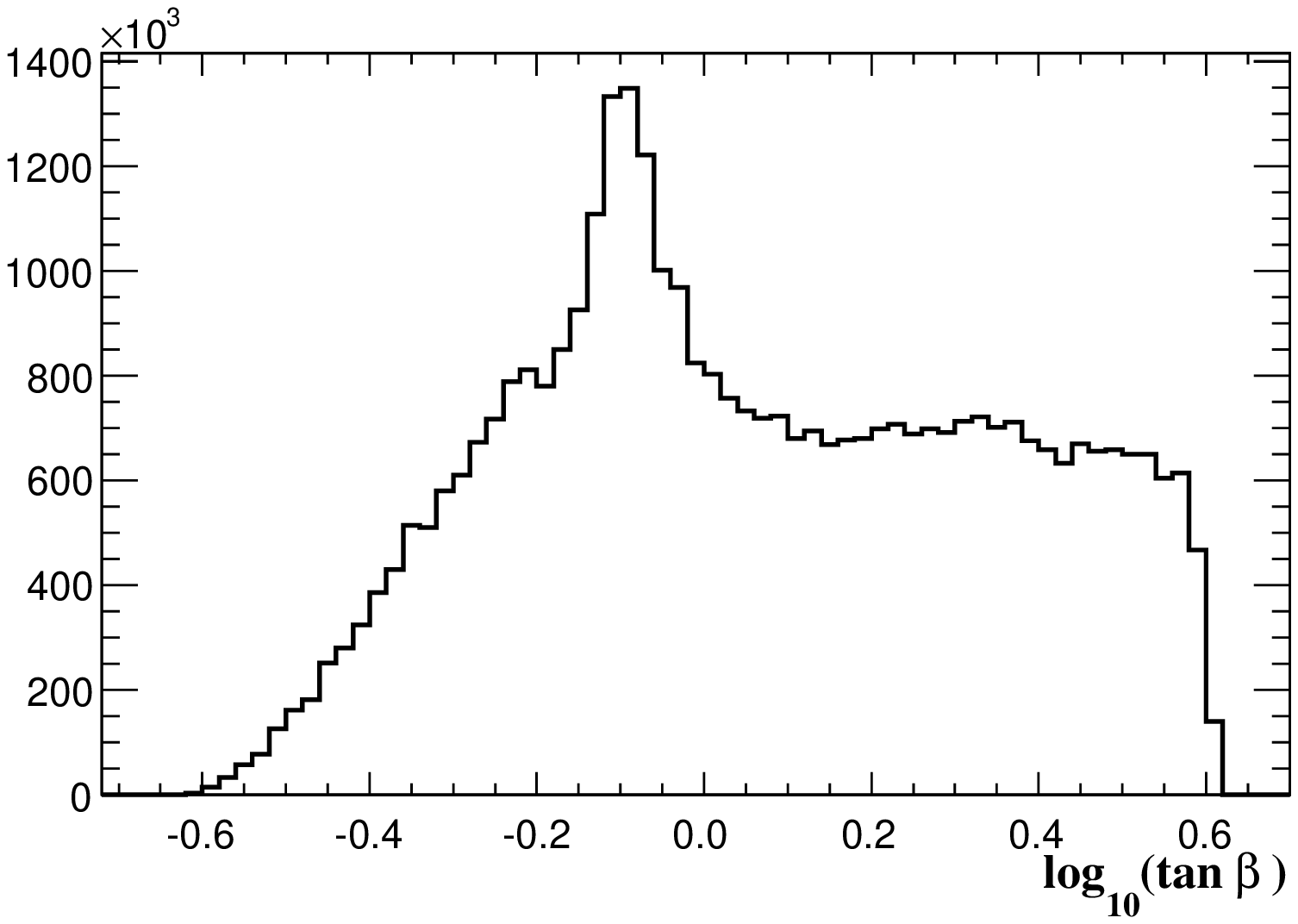}
\caption{Left: Contour plots of the probability densities in 2HD4G with
the mass of the lightest ${\mathcal CP-}$even state $m(h) = 124.5$ GeV.
Quark mass splitting \vs~lepton mass splitting. The box marks the area
where the magnitude of the mass splittings is less than $M_W$.  All scales
are in $\gev$; probability densities have been normalized and each bin is
$10\gev \times 10\gev$.  Right: The probability density function for
$\log_{10}(\tan \beta)$~in 2HD4G with the mass of the lightest
${\mathcal CP-}$even state $m(h) = 124.5$ GeV.}
\label{fig:THDM_dM_tanB}
\end{center}
\vspace{0.2em}
\end{figure*}

\begin{figure*}[t]
\begin{center}
\includegraphics[width=\textwidth]{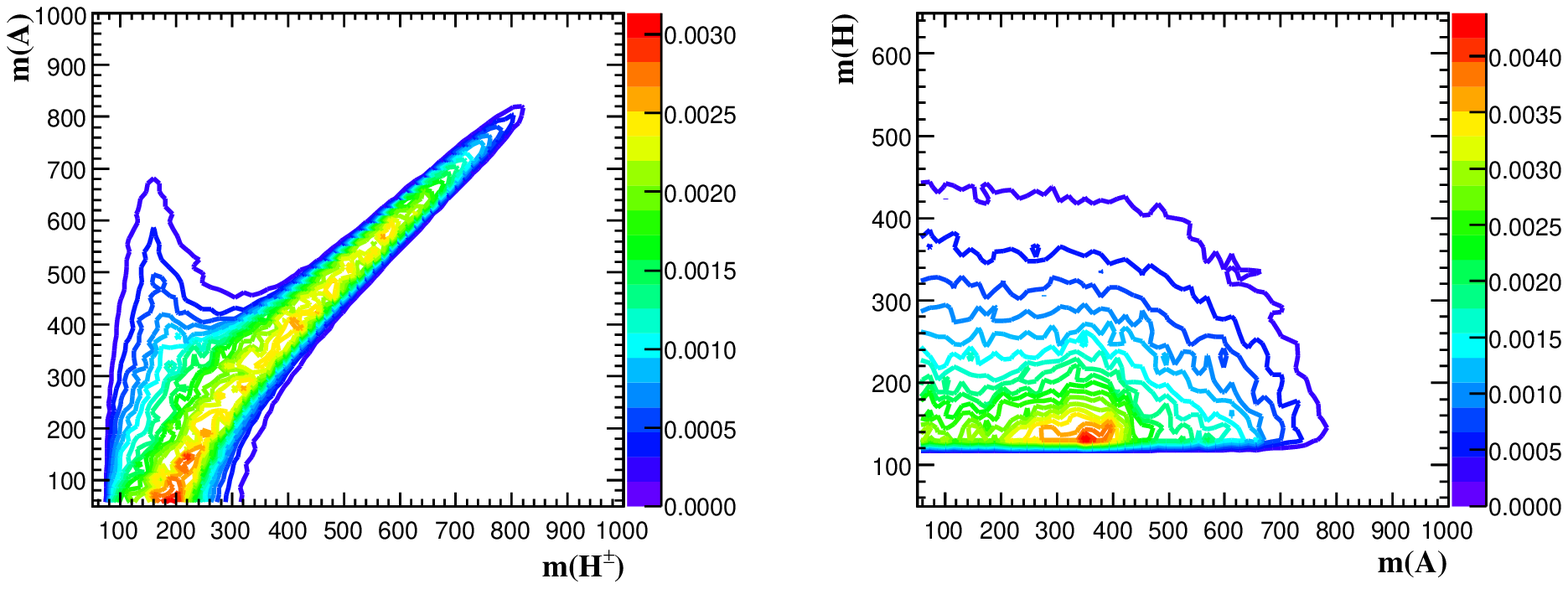}
\caption{The probability density function for the masses of $H$, $A$ and
$H^\pm$ in 2HD4G with the mass of the lightest
${\mathcal CP-}$even state $m(h) = 124.5$ GeV.}
\label{fig:THDM_mH}
\end{center}
\vspace{0.2em}
\end{figure*}

Carpenter and Rajaraman~\cite{bib:RHnu4} revisited the LEP II results in a
scenario with both left- and right-handed neutrinos.  They conclude that
\mN as low as $62.1\gev$ is possible.  Some recent
studies~\cite{bib:GoodInj,bib:CloseCall,bib:gamgamOK2} also consider
low values of $m(\nu_4)$. We find that lowering the bound on \mN to $62.1\gev$
does not produce much change relative to our baseline scenario.
Figure~\ref{fig:RHn4_M12} shows distributions that have the same probabilities
of mass splittings less than $M_W$ and the same probabilities of normal mass
hierarchies as our baseline scenario to within about 2\%.


\section{\label{sec:doubles}TWO HIGGS DOUBLET SCENARIO}

\subsection{Method}

Should the hints of a Higgs boson with $m(h) = 124.5$ GeV solidify with
more data, the 4G hypothesis is only tenable if an extended electroweak
symmetry breaking sector exists.  As an example of such an extension
we consider a second Higgs doublet~\cite{bib:Branco} in conjunction with
a fourth sequential fermion generation (2HD4G).  Two identical complex
scalar $SU(2)_L$ doublet fields $\Phi_1$ and $\Phi_2$, both of hypercharge
$Y= 1$ are postulated.  To forbid flavor changing neutral currents, we
select the Type II Yukawa coupling pattern, in which $Q = 2/3$~ quarks
couple to one doublet and $Q = -1/3$~ quarks and charged leptons to the other.
This restriction permits a $\mathbb{Z}_2$~symmetry to distinguish $\Phi_1$~from $\Phi_2$.
We restrict consideration to the gauge invariant, renormalizable and
${\mathcal CP-}$conserving potential
\begin{align}
V  & =m_{11}^{2}\Phi_{1}^{\dagger}\Phi_{1}+m_{22}^{2}\Phi_{2}^{\dagger}%
\Phi_{2}+\frac{\lambda_{1}}{2}(\Phi_{1}^{\dagger}\Phi_{1})^{2}+\frac
{\lambda_{2}}{2}(\Phi_{2}^{\dagger}\Phi_{2})^{2}	\nonumber\\
& +\lambda_{3}(\Phi_{1}^{\dagger}\Phi_{1})(\Phi_{2}^{\dagger}\Phi_{2}%
)+\lambda_{4}(\Phi_{1}^{\dagger}\Phi_{2})(\Phi_{2}^{\dagger}\Phi_{1})	\nonumber\\
& +\frac{\lambda_{5}}{2}((\Phi_{1}^{\dagger}\Phi_{2})^{2}+(\Phi_{2}^{\dagger}\Phi_{1})^{2})
\end{align}
where all the parameters $m_{ii}$ and $\lambda_i$ are real. This system and
its vacua preserve an additional $\mathbb{Z}_{2}$ symmetry. There are
two ${\mathcal CP-}$even neutral bosons, $h$ and $H$ ($m(h) < m(H)$) a
${\mathcal CP-}$odd neutral boson, $A$, and the charged bosons $H^\pm$
in this model.

This model is different from the similarly-named ``4G2HDM'' model
of~\cite{bib:differs}; however,~\cite{bib:RGE_2HD4G} analyzed a similar
model prior to the appearance of the $124.5$ GeV hint. The
presence of large fourth generation Yukawas can lead to large radiative
corrections which can potentially destabilize the form of the Higgs potential.
Here we assume that the 2HDM effective potential is stabilized by some effect
near the TeV scale, so that we can focus on the resulting effective theory
below the TeV scale.

Though it is beyond the scope of this study, the combination of two Higgs
doublets with a fourth sequential fermion
generation creates a rich phenomenology for which constraints from the
kaon and $b$~sector could be derived.  For example, the coupling constants
$Z \rightarrow$ \bbb~vertex will obtain corrections which depend on $V_{t_4b}$,
$m(t_4)$, and (depending on chirality) either \tanB~or its inverse; these
contributions can be constrained experimentally.

Two important parameters of this model are \tanB, the ratio of the
vacuum expectation values of the two doublets and $\alpha$, the
angle which diagonalizes the mass-squared matrix of the ${\mathcal CP-}$even
bosons.  Values of $\tan \beta$ less than 1 are disfavored experimentally
assuming three fermion generations; more generally, $\tan \beta > 0.3$ results
from the requirement that the top quark Yukawa coupling not exceed the
perturbative limit~\cite{bib:Branco}. Requiring perturbativity of the fourth
generation Yukawa interactions can impose additional constraints. For the sake
of generality, however, we do not impose this additional restriction in our
scans.  We sample $\tan \beta$ in a scale-independent way, \ie, the distribution of
$\log($\tanB$)$ is uniform.  The angle $\alpha$~is scanned uniformly but
samples are weighted according to the value of $\alpha$~as described below;
the masses of the 4G fermions and the bosons $H$, $A$ and $H^\pm$ are selected
with an initially uniform distribution.  The mass of the lightest
${\mathcal CP-}$even boson is set to $124.5\gev$.  For further discussion on
the phenomenology of two Higgs doublets with a fourth fermion generation,
including the case where $\nu_4$~is stable and contributes invisible decays
to either $h$~or $A$, see~\cite{bib:ChenHe}.

The $124.5\gev$ hint is strongest in the channel
$gg \rightarrow h \rightarrow \gamma\gamma$, where the ATLAS experiment reports
an excess above background of 2.8 standard deviations.  The second most 
significant hints are in the channels $gg \rightarrow h \rightarrow VV^*$,
where the ATLAS results have a significance of 2.1 and 1.4 standard 
deviations for $V = W,Z$~ respectively.  The combination of ATLAS and CMS data
correspond to a $\gamma\gamma$~production rate about $1.4\pm0.7$ times the
prediction of the Standard Model~\cite{bib:Carmi2012}; for $VV^*$, it is about 
0.8\pma{0.7}{0.4}~ times the Standard model rate.  

For each scanned value of $\alpha$, we calculate
\begin{center}
$\sigma(gg \to h) \frac{\Gamma(h \to \gamma\gamma  )}
					   {\Gamma_{2HD4G}(tot)}$
and
$\sigma(gg \to h) \frac{\Gamma(h \to VV^*  )}
					   {\Gamma_{2HD4G}(tot)}$
					   \end{center}
for the 2HD4G scenario and form a $\chi^2$~of these values against these experimental values.  We weight
each sample according to that $\chi^2$.  We do not consider constraints
from decays of the $H$, $A$ and $H^\pm$~ which are very parameter dependent
in the 2HD4G scenario.

The dominant production mechanism at both the
LHC and the Tevatron is gluon fusion through loop diagrams involving the
colored states. In a 2HDM, this includes the contributions from both the $t$,
$b$ as well as $t_4$ and $b_4$. The Standard Model normalized cross section
$\widehat{\sigma} = \sigma_{2HD4G} / \sigma_{SM}$ of the gluon fusion
production cross sections is:
\beq
\widehat{\sigma} =\frac{
	|\frac{c_{\alpha}}{s_{\beta}}(A_{1/2}(t) +A_{1/2}(t_4))
	-\frac{s_{\alpha}}{c_{\beta}}(A_{1/2}(b) + A_{1/2}(b_4))|^{2}
}{|A_{1/2}(t)|^{2}}
\eeq
where $c_{\alpha}$ and $s_{\alpha}$ denote $\cos \alpha$ and $\sin \alpha$
respectively, and $A_{1/2}(X)$ is the threshold correction of a spin $1/2$
particle $X$ to the $h \rightarrow gg$ vertex for a $124.5$ GeV Higgs, with
notation as in \cite{bib:tHHG}. A similar expression holds for the Standard
Model normalized decay rate $h \rightarrow \gamma \gamma$. In a 2HDM, this will
include terms from loops containing $W$, $t$, $b$ and $\tau$ and charged fourth
generation fermions, as well as a contribution from $H^\pm$,
which all depend on the mixing angles. The total width of
the Higgs in 2HD4G including the mixing angle dependence is fixed by similar
considerations. Much as in \cite{bib:Heckman:2012nt}, the overall normalization
can be extracted from the recently updated values for the Standard Model
$124.5$ GeV Higgs partial widths \cite{bib:Barger:2012hv} by including the
mixing angle dependence and contribution from extra states in the various
2HD4G partial widths.


Constraints on two doublet models are readily available~\cite{bib:THDMC}
through the package \MC.  We observe the constraints of tree-level
unitarity~\cite{bib:unitMC}, perturbativity (\ie, the magnitudes of all
the quartic Higgs couplings must be less than $4\pi$), and the absence of
runaway directions, as implemented in \MC.  Contributions to the oblique
electroweak parameters~\cite{bib:obieMC} are also provided as part of \MC.

\subsection{Results}

Figures~\ref{fig:THDM_M12} and~\ref{fig:THDM_dM_tanB} show the lepton and
quark mass spectra in our two Higgs doublet scenario.  The quark (lepton)
mass splittings are less than $M_W$ in 99\% (65\%) of our samples;
normal mass hierarchies in the quark (lepton) sector occur with a probability
of 59\% (72\%).

Low values of \tanB~are likely in 2HD4G; in
Figure~\ref{fig:THDM_dM_tanB}, \tanB~$< 1$~in 46\% of the final probability
density function.  Figure~\ref{fig:THDM_mH} shows the distribution of Higgs
boson masses.  There is a strong correlation between the masses $m(H^\pm)$~
and $m(A)$~largely but not entirely created by requiring $VV^*$~as well as
$\gamma\gamma$~production to be in agreement with experiment.
It is amusing to note that the most likely values for the mass of the
second ${\mathcal CP-}$even boson are just over $124.5\gev$, and masses
corresponding to a small excess in the $4\ell$ channel at $240\gev$ are not
improbable.

While the extended Higgs sector does alter the results from the single
Higgs double scenarios, the broad features of the mass splitting structures
and preference for low masses, particularly for $\nu_4$, remain.  These
features are largely a result of the structure of the contributions to
the electroweak oblique parameters from the fourth generation of sequential
fermions.  Similar results might be expected in almost any extension to the
Higgs sector that is broadly consistent with a Standard Model-like Higgs.

\section{\label{sec:sum}SUMMARY}

While stringent limits on \mT~and \mB~have been found in specific
decay modes by the LHC, completely ruling out the fourth generation
hypothesis requires an analysis~\cite{bib:Irvine} that combines the results
from a number of modes to obtain a result that is independent of quark
mixing in the fourth generation.

We have used sampling methods to determine the probability densities
of the masses of a possible fourth sequential generation of fermions
in scenarios with one or two Higgs doublets. With a single Higgs
doublet and a low ($115$ or $124.5\gev$) Higgs mass, fourth generation
mass splitting in the quark sector is less than
$M_W$~ (see also~\cite{bib:GoodInj}).  Quark sector mass splittings less
than $M_W$ are favored but less certain if the Higgs mass is $600\gev$.
A fourth generation is on the verge of being ruled out in the case of
a single Higgs doublet~\cite{bib:LHC_Higgs4}, but a Type II two Higgs
doublet model can be designed to reproduce the hints at $124.5\gev$ from
the LHC and the Tevatron.  In that case, quark mass splittings less than
$M_W$ are still favored.  In all of our scenarios, the most favored values
for $m(\tau_4)$ are just above the experimental limit of $110.8$ GeV,
making searches for a fourth generation charged lepton an interesting
possibility.

\section{\label{sec:thanks}ACKNOWLEDGEMENTS}

We gratefully thank the authors of \MC \,\, in particular for developing a
new version of their code. We also
thank P. Kumar and P. Langacker for helpful discussions. J.J.H thanks the Simons Center for Geometry and Physics 
for hospitality during the completion of this work. The work of L.B. and E.R.-H. is supported by DOE contract
DE-AC02-07CH11359; the work of J.E. is supported by CONACyT (M\'exico)
contracts 82291--F and 151234; the work of J.J.H. is supported by NSF
grant PHY-0969448 and by the William Loughlin membership at the
Institute for Advanced Study.


\end{document}